\documentclass{workshop}
\begin{document}

\title{
The Gamow Explorer: \\ A Gamma-Ray Burst Mission to Study the High Redshift Universe} 

\author{
Nicholas E. White,$^1$
\\[12pt]  
%
$^1$  George Washington University, Department of Physics,
715 21st Street NW, Washington DC 20052  \\
%
\textit{E-mail: newhite@gwu.edu} 
}

\abst{
Long Gamma Ray Bursts (LGRBs) can be used to address key questions on the formation of the modern universe including: How does the star formation rate evolve at high redshift? When and how did the intergalactic medium become re-ionized? What processes governed its early chemical enrichment? A LGRB signals when a massive star collapses to form a black hole and in doing so provides an independent tracer of the star formation rate. The LGRB afterglow is a bright back-light to view the host galaxy and intergalactic medium in absorption. The \emph{Gamow Explorer} will be optimized to search for high redshift LGRBs, with a \emph{z}$>$6 detection rate at least ten times \emph{Swift}. Furthermore it will go beyond \emph{Swift} by using the photo-\emph{z} technique to autonomously identify $>$80\% of \emph{z}$>$6 redshift LGRBs to enable rapid follow up by large ground based telescopes and JWST for spectroscopy and host galaxy identification. The \emph{Gamow Explorer} will be proposed to the 2021 NASA MIDEX opportunity for launch in 2028.
}

\kword{Workshop: Proceedings; Gamma Rays: Bursts; Early Universe; Astrophysics: High Energy Astrophysical Phenomena; Space Vehicles: Instruments; Telescopes; Astrophysics}

\maketitle
\thispagestyle{empty}

\section{Introduction}

Long Gamma Ray Bursts (LGRBs) have been observed out to very high redshift (\emph{z}), with currently the most distant at \emph{z}$\sim$9.4 \citep{2011ApJ...736..7S}. LGRBs provide both a tracer of massive star formation and a bright back light to determine the properties of the intervening material in absorption from the host galaxy and the intergalactic medium.  After the initial Gamma-ray burst there is a multi-wavelength afterglow that fades over hours to a few weeks. The afterglow is the result of a relativistic jet from the newly formed black hole plowing into the surrounding interstellar medium. The early afterglow K band brightness can range from apparent magnitude 16 to 22 for up to several hours to even days after the trigger. As such LGBRs can be used by large telescopes as probes of the high redshift Universe including the epoch of reionization, early star formation and the first chemical enrichment of the intergalactic medium. 

The \emph{Neil Gehrels Swift Explorer} \citep{2004ApJ...611..1005G}, launched in 2005, was designed as a multi-wavelength observatory to both detect GRBs and rapidly point X-ray and UV-Optical telescopes to accurately determine the position and identify the host galaxy. This was a highly successful strategy and provided a detailed measurement of the LGRB redshift distribution, including finding a handful of LGRBs above \emph{z} of 6. However, ground based telescopes are required to determine the redshift and approximately only one third of \emph{Swift} LGRBs have measured redshifts. A total of $\sim$6 have been found at high redshift (\emph{z}$>$6) over the 15 yr \emph{Swift} mission life to date. 

The \emph{Gamow Explorer} is designed to take the next step by: 1) maximizing the number of high redshift LGRBs (\emph{z}$>$6) detected per year, with at least an order of magnitude more than \emph{Swift}; 2) autonomously identify \emph{z}$>$6 LGRBs and measure their redshift; and 3) having a field of regard to maximize follow up by ground and space based telescopes, including NASA's James Webb Space Telescope (JWST).

The \emph{Gamow Explorer} is named in honor of George Gamow for his seminal contribution to the Big Bang theory of the expanding Universe. It will be proposed in late 2021 to the next NASA Medium Class Explorer (MIDEX) opportunity. If approved by NASA the \emph{Gamow Explorer} would launch in 2028. This timing is critical to ensure overlap with JWST during its 10 year design lifetime. 

\subsection{Star Formation at High Redshift}

LGRBs trace the core-collapse of massive stars and as such represent an independent direct tracer of the star formation rate (SFR). \citet{2012ApJ...744..95R} derived the SFR rate from a sample of LGRBs and found at \emph{z}$>$4 it declines more slowly than the SFR rate based on galaxy observations. Possible explanations are that the fraction of LGRB increases with redshift or there is an evolution in the luminosity of GRBs with redshift. Another factor is that LGRBs seem to favor lower metalicity environments \citep{2019A&A...623A..26P}. LGRBs provide a possible means to observe the first population of stars that formed, the so called Pop III stars. These massive stars, likely hundreds of solar masses, will form in a metal free environment and live for a relatively short time and to be prominent at z $\sim$ 10 to 20 \citep{2003ApJ...596..34S}. Pop III stars that undergo core-collapse may produce LGRBs from jets driven by magneto-hydrodynamic processes that scale with the mass of the star \citep{2011ApJ...731..127T}. LGRBs may be the only way to directly observe a Pop III star, by observing the end of its life.

\subsection{The epoch of re-ionization}

The epoch of Hydrogen reionization occurred between redshift 6 and 15 with the possible ionizing sources: massive stars, quasars, X-ray binaries, and dark matter decays. Reionization models that invoke UV radiation from massive stars require a high proportion of the stellar ionizing radiation escapes the host galaxy. The Ly$\alpha$ absorption feature in LGRB afterglows can be used to constrain the optical depth and topology of reionization \citep{2019MNRAS...483..5380S}. LGRBs are ideal probes to investigate this issue, as their massive progenitors reside in the same location. Their high luminosity at early times and featureless power-law spectra make them an ideal back-light to study the intervening material. \citet{2019MNRAS...483..5380S} compiled LGRB optical depth measurements to date from lower redshifts and the results so far show high optical depths. This sample needs to be extended to the high redshift universe. 

\subsection{Chemical Enrichment from the first stars}

GRBs are ideal tools for probing the metal enrichment in the early IGM, due to their high luminosity and featureless power-law spectra. Metal absorption lines are imprinted according to the Pop III SN event: Pair Instability vs. core-collapse \citep{2012ApJ...760...27W}. High resolution spectroscopy  of the LGRB afterglow a few hours or days after the LGRB trigger by the coming thirty meter class ground based telescopes will be able to distinguish whether the first heavy elements were produced in a pair-instability or a core-collapse supernova, thereby constraining the initial mass function of the first stars. 

\section{The Gamow Explorer Mission Concept}

To create a sample of high redshift LGRBs requires to autonomously and promptly determine which of the hundreds of LGRBs are at \emph{z}$>$6. The photo-\emph{z} technique is well proven for rapidly identifying high redshift objects. It takes advantage of the Hydrogen Lyman $\alpha$ absorption which creates a sharp blue-ward drop out. High redshift LGRBs from \emph{Swift} were identified using the photo-\emph{z} technique from the ground e.g. GRB 090423 at z $\sim$ 8.2 \citep{2009Nature...461..1254S,2009Natur.461.1258S}. To make further progress requires that the redshift determination be done autonomously as part of the observatory. 

For the \emph{Gamow Explorer} two instruments are required: one to detect LGRBs with an arc minute position and the other an IR telescope to precisely determine the position and the redshift. As with \emph{Swift} a rapidly slewing spacecraft uses the GRB arc minute position to point an IR telescope at the GRB location. Onboard software identifies the Lyman drop out and then broadcasts an alert with the redshift determination, along with the LGRB time, an accurate arc second celestial location and IR flux. The mission configuration and predicted LGRB rates given here are preliminary and will continue to be refined to meet the science goals.




\subsection{Burst Event Telescope for Alerts}

A key requirement is to maximize the number of high redshift LGRBs. \citet{2015MNRAS...448..2514S} has shown that this is achieved by observing in the X-ray band. This is primarily because at high redshifts the spectral peak of the LGRB emission is shifted to lower energies due to the Cosmological redshift and the corresponding flux diluted over a large distance. We have undertaken a study of two approaches: a Lobster Eye X-ray Telescope (LEXT) and an X-ray Coded-mask Telescope (XCAT).

The LEXT is an array of slumped micro channel plates on a spherical frame \citep{2002SPIE...4497..115F}. Each module has a 30$\times$30 deg field of view (FOV) with a core point spread function of 6.5 arc min giving $\sim$1 arc min locations. LEXT covers the 0.5 to 5 keV band. There are two or three modules covering 0.5 to 0.75 sr. Each module has a $15\times15$cm focal plane with either nine $5\times5$cm CCDs or sixteen $3.7\times3.7$cm CMOS chips. 

The XCAT is composed of 15 modules each with a FOV of $20\times20$ deg fully coded \citep{2010SPIE.7732E..4FF}. The open grasp is 15.3 cm$^2$ sr per module. They cover the 2 to 10 keV energy band, with localization of $\sim$0.5 arc min. Each module has four CMOS chips. 

Predicted LGRB rates for \emph{Gamow Explorer} have been obtained for both cases using the modeling of \citet{2015MNRAS...448..2514S}. The LGRB rates predicted for \emph{Swift} were also computed, to ensure consistency with the observations. The predicted rates are shown in Table 1. The uncertainties are highlighted in Table 1 for the \emph{z}$>$6 case. The \emph{z}$>$6 rate is 10 times the \emph{Swift} rate, This increases to a factor of $\sim$ 30 if redshift identifications are included.

Because it is a focusing system the LEXT provides an order of magnitude more sensitivity and is better aligned to detect cosmological time dilated and dimmed high redshift GRBs. Based on this and other details of the trade study e.g. cost, the LEXT approach has been adopted as the baseline.  

\begin{table}[h]
\caption{\emph{Gamow Explorer} LGRB preliminary predicted detection rates for a high 95\% efficiency orbit following \citet{2015MNRAS...448..2514S}}
\begin{center}
\begin{tabular}{lccc} \hline\hline\\[-6pt]
Redshift        & LEXT3& XCAT15 & \emph{Swift}\\ 
       & yr$^{-1}$ & yr$^{-1}$ & yr$^{-1}$ \\ \hline
 \emph{z} $>$ 0 & 314 &       345 &  77 \\
 \emph{z} $>$ 5   & 26 &       19  & 3\\
 \emph{z} $>$ 6   & 15 (9-26) &       11 (6-19) & 1.4 \\
 \emph{z} $>$ 7   & 10 &       7  & 0.8\\  
 \emph{z} $>$ 10  & 3 & 2 & 0.2\\ \hline 
\end{tabular}
\end{center}
\end{table}

\subsection{Accurate Location and Photo-\emph{z} Alerts}

From redshift 6 to 20 the Lyman drop out crosses the 0.6 to 2.5 micron band. The required sensitivity is calculated based on observed high redshift LGRB afterglows and transforming lower redshift afterglows to redshift 10 \citep{2010ApJ...720.1513K}. Given the initial rapid decay of the afterglow a 500s exposure is required within 100s of the LGRB trigger. We find that a limiting sensitivity of $15\mu$Jy will retrieve at least 80\% of the high redshift afterglows. An aluminum three-mirror anastigmat telescope design provides the required 10 arc min FOV. It is passively cooled to 200K. Four channels are obtained by using dichroic mirrors to image onto a single detector. The required sensitivity can be achieved with a 30cm diameter. A slit spectrometer with R$\sim$20 obtains a follow up spectrum to confirm the Lyman break and redshift in a 6000s exposure. The focal plane is a flight qualified spare JWST NIRCam detector, sidecar electronics and focal plane assembly. These H2RG detectors have $2048\times2048$ pixels, cover the 0.6 to 2.5 micron band and passively cooled to 100K. 

\subsection{Orbit}

We have studied both L2 and Sun-Sync orbits. Based on a JPL Team-X study we conclude that only an L2 halo orbit meets the requirements. Low Earth orbits require a stare-step strategy to maintain the IR telescope Earth limb constraint. They are only 72\% efficient and do not allow year round visibility of the JWST field of regard. The L2 orbit meets all the science requirements, simple operations and high efficiency. The communications from L2 of the alerts is via a continuous low bit rate (500 bps) link, based on the capabilities in use for space weather missions at L1 \citep{1998SSRv...86..633Z}. 

\subsection{Concept of Operations}

The basic concept of operations is to monitor a patch of the sky to wait for a LGRB. When it is detected the arc min position is used to autonomously slew the spacecraft to point the IR telescope at its location within 60s. A bright/variable IR counterpart is identified and an arc sec location, flux and a photo-\emph{z} determination obtained within a 500s exposure. Parameters are transmitted to ground as soon as they are available. If the LGRB is at \emph{z}$>$6 then JWST and ground based telescopes observations are triggered. The \emph{Gamow} observations continue with a trim maneuver to place the IR counterpart on the slit of a low resolution (R$\sim$20) spectrometer to confirm the photo-\emph{z} redshift.

\section{Acknowledgments}

\vspace{1pc}
\noindent This paper represents the work of the entire \emph{Gamow Explorer} team from: George Washington University, University of Leicester, Penn State University, MIT Kavli, JPL, LANL, SWRI, GSFC, MSFC, Clemson University, Harvard University, STScI, Radboud University, INAF-OAB, and IAA. I want to recognize the contributions of M. Bautz, E. Berger, A. Falcone, C. Feldman, D. Fox, A. Fruchter, G. Ghirlanda, S. Guiriec, A. van der Horst, D.A. Kann, C. Kouveliotou, C. Lawrence, A. Lien, A. Nash, P. O'Brien, D. Palmer, P. Roming, R. Salvaterra, R. Willingale, C. Wilson-Hodge, and E. Young. NASA JPL provided a Team-X mission design study.

\label{last}

\end{document}